\PassOptionsToPackage{hidelinks}{hyperref}
\documentclass{article}
\usepackage{spconf,amsmath,amssymb,graphicx,hyperref,booktabs,multirow}
\def\X{{\mathbf X}}
\def\V{{\mathbf V}}
\def\A{{\mathbf A}}
\def\L{{\cal L}}

\title{{GIVE-KWS}: Gated Injection of Visual Evidence for\\
       Noise-Robust Query-by-Example Keyword Spotting}

\name{\href{https://orcid.org/0009-0004-6936-2576}{Ming-Hsiang Hu}$^{1}$,
      \href{https://orcid.org/0009-0003-1417-9638}{Kuan-Tang Huang}$^{1}$,
      \href{https://orcid.org/0000-0001-7044-9434}{Hung-Shin Lee}$^{2}$,
      \href{https://orcid.org/0000-0003-0693-8932}{Berlin Chen}$^{1}$}
\address{$^{1}$Department of Computer Science and Information Engineering, National Taiwan Normal University\\ $^{2}$Graduate Institute of AI Interdisciplinary Applied Technology, National Taiwan Normal University}

\begin{document}
\ninept
\maketitle

\begin{abstract}
Visual speech promises noise-robust keyword spotting, yet a visual stream is not necessarily used. On a tri-modal query-by-example keyword spotting (QbyE-KWS) benchmark, we find that a system with a task-trained visual encoder comes within 2 percentage points of a text-and-audio system in equal error rate (EER) at -10~dB, and link this gap to the encoder's lack of phonemic information. We present GIVE-KWS, whose fusion stage, GIVE (Gated Injection of Visual Evidence), conditions query audio on lip motion through gated cross-attention. We show that visual robustness depends on two interacting conditions: a phoneme-bearing visual representation, and fusion that injects visual evidence rather than rescaling audio features. Under a phoneme-bearing encoder, injection yields an effective SNR gain of 4.0--9.3~dB over masking at -10~dB, whereas under a phoneme-poor one it nearly vanishes. Relative to the benchmark system, GIVE-KWS reduces unseen-keyword EER by 72.9\% at -10~dB and 62.8\% on average.
\end{abstract}

\section{Introduction}
\label{sec:intro}

Keyword spotting (KWS) detects user-specified terms in continuous speech and underpins voice interfaces, media retrieval and spoken content search. Speech systems that rely on audio alone degrade sharply under realistic acoustic conditions---background noise, reverberation, far-field attenuation and overlapping talkers all erode the acoustic evidence on which the decision rests~\cite{cai2020,lim2024,wang2025}. Visual speech offers a channel that is unaffected by acoustic corruption and that carries complementary articulatory information: lip shape constrains the place of articulation even when the corresponding acoustic cues have been buried. This complementarity motivates audio-visual KWS~\cite{wang2024,zhang2023,xiong2025}.

We study query-by-example (QbyE) KWS, in which a new keyword is specified at test time by spoken enrollment examples rather than fixed in advance (as in traditional KWS~\cite{chen2015}), and the system decides whether a query contains the same keyword. User-defined KWS systems based on phoneme-level matching~\cite{kreuk2023}, multimodal prompts~\cite{ai2024}, contrastive alignment~\cite{jung2025}, two-stage verification~\cite{ai2026} and speaker conditioning~\cite{hu2026} all operate on audio alone. Matching is fragile in a specific way: unlike recognize-then-match pipelines, it has no lexical or language-model prior to fall back on, so noise that corrupts the query representation enters the similarity computation unmediated.

\begin{figure*}[t!]
\centering
\includegraphics[width=\textwidth]{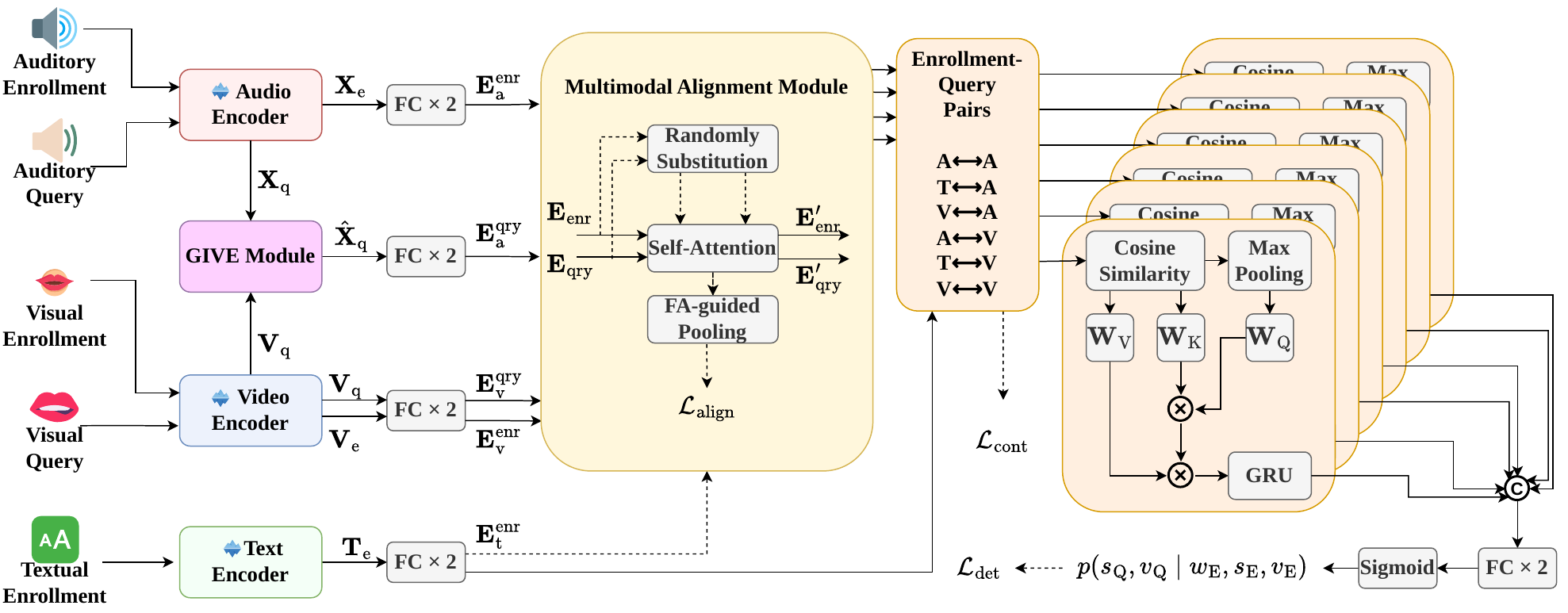}\\[-5pt]
{\small (a) Overall framework.}
\vspace{-10pt}
\end{figure*}

\begin{figure}[t]
\centering
\includegraphics[width=\columnwidth]{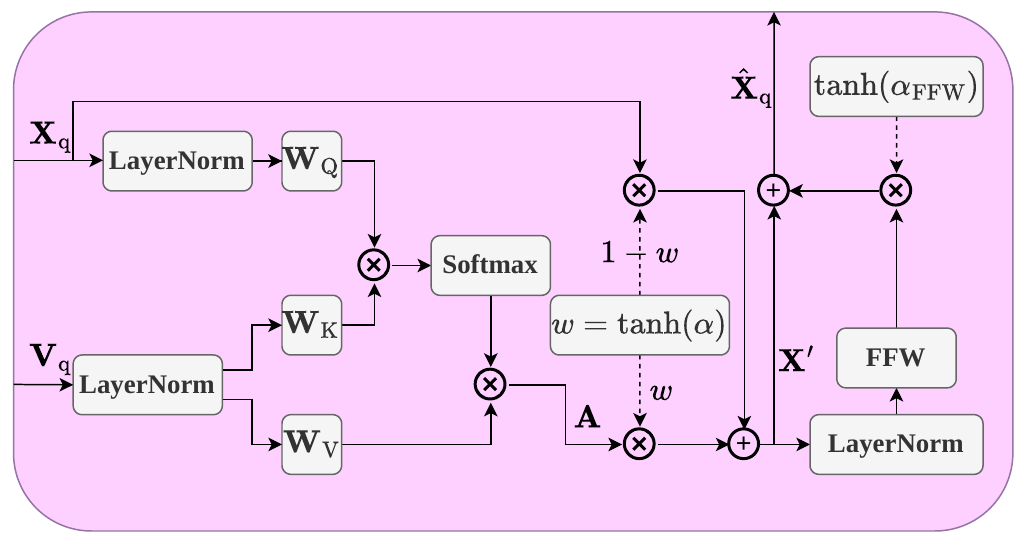}\\[-1pt]
{\small (b) GIVE module.}\\

\caption{{GIVE-KWS}. Frozen encoders produce text, audio and visual representations; the query audio is conditioned on lip motion by the GIVE module before projection; the alignment stage anchors audio and video to text phoneme embeddings; six enrollment--query similarity branches are combined into a detection score. Dashed arrows denote training objectives. In the GIVE module, query audio frames attend over the visual sequence, and the attended evidence $\A$ is blended into the audio stream with complementary weights $1-w$ and $w$.}
\label{fig:arch}
\vspace{-10pt}
\end{figure}

\begin{table*}[t]
\centering
\caption{Results on MISP-QEKS (\%). Avg.\ is the mean over all five conditions. The \emph{Text\,+\,audio only} system is trained without the visual stream; rows sharing a visual encoder differ only in fusion: masking (XEQ-Matcher) or injection (GIVE-KWS). All systems are retrained under our protocol; best upper-block results in bold. Relative gains are AUC increases and EER reductions; \emph{Overall} compares GIVE-KWS with XEQ-Matcher using its original encoder; the \emph{Fusion} rows compare injection with masking under the same encoder. $^{*}$: significant (paired bootstrap, Bonferroni over six comparisons); for the CNN-ResNet fusion row at $-10$~dB the $95\%$ interval includes zero on both splits.}
\vspace{3pt}
\label{tab:main}
\small
\setlength{\tabcolsep}{2.8pt}
\begin{tabular}{llcccccccccccc}
\toprule
\multirow{3}{*}{System} & \multirow{3}{*}{Visual enc.} &
\multicolumn{6}{c}{Eval-seen} &
\multicolumn{6}{c}{Eval-blind} \\
\cmidrule(lr){3-8}\cmidrule(lr){9-14}
 & & \multicolumn{2}{c}{Clean} & \multicolumn{2}{c}{Avg.} & \multicolumn{2}{c}{$-10$~dB}
   & \multicolumn{2}{c}{Clean} & \multicolumn{2}{c}{Avg.} & \multicolumn{2}{c}{$-10$~dB} \\
 & & AUC & EER & AUC & EER & AUC & EER & AUC & EER & AUC & EER & AUC & EER \\
\midrule
Text\,+\,audio only          & ---        & 94.97 & 13.20 & 86.23 & 22.43 & 73.93 & 33.53 & 97.99 & 7.92 & 87.98 & 19.71 & 71.72 & 35.15 \\
\midrule
XEQ-Matcher~\cite{xiong2025}  & CNN-ResNet~\cite{zhou2022} & 95.00 & 13.37 & 86.33 & 22.39 & 74.11 & 33.52 & 98.20 & 7.37 & 89.17 & 18.46 & 74.15 & 33.19 \\
GIVE-KWS                     & CNN-ResNet~\cite{zhou2022} & 95.21 & 13.08 & 86.70 & 22.10 & 74.45 & 33.18 & 98.44 & 6.80 & 89.49 & 17.97 & 74.21 & 33.13 \\
XEQ-Matcher~\cite{xiong2025}  & AV-HuBERT~\cite{shi2022a,shi2022b}  & 96.61 & 10.09 & 93.73 & 14.07 & 90.05 & 18.67 & \textbf{98.96} & 5.25 & 97.88 & 7.43 & 96.40 & 10.14 \\
\textbf{GIVE-KWS}            & AV-HuBERT~\cite{shi2022a,shi2022b}  & \textbf{97.53} & \textbf{7.92} & \textbf{95.81} & \textbf{10.75} & \textbf{93.62} & \textbf{13.96} & 98.94 & \textbf{4.91} & \textbf{97.99} & \textbf{6.87} & \textbf{96.72} & \textbf{9.01} \\
\midrule
\multicolumn{14}{l}{\textit{Relative gain (\%)}} \\
\multicolumn{2}{l}{\quad Overall}
  & 2.7\% & 40.8\% & 11.0\% & 52.0\% & 26.3\% & 58.4\%
  & 0.8\% & 33.4\% & 9.9\% & 62.8\% & 30.4\% & 72.9\% \\
\multicolumn{2}{l}{\quad Fusion (AV-HuBERT~\cite{shi2022a,shi2022b})}
  & 1.0\% & 21.5\%$^{*}$ & 2.2\% & 23.6\%$^{*}$ & 4.0\% & 25.2\%$^{*}$
  & 0.0\% & 6.5\% & 0.1\% & 7.5\%$^{*}$ & 0.3\% & 11.1\%$^{*}$ \\
\multicolumn{2}{l}{\quad Fusion (CNN-ResNet~\cite{zhou2022})}
  & 0.2\% & 2.2\% & 0.4\% & 1.3\% & 0.5\% & 1.0\%
  & 0.2\% & 7.7\% & 0.4\% & 2.7\% & 0.1\% & 0.2\% \\
\midrule
\multicolumn{14}{l}{\textit{GIVE-KWS (AV-HuBERT~\cite{shi2022a,shi2022b}), visual stream at inference}} \\
\quad removed                &            & 95.88 & 11.32 & 87.56 & 20.36 & 74.79 & 32.21 & 98.03 & 6.62 & 89.04 & 17.80 & 73.87 & 32.91 \\
\quad zeroed                 &            & 95.70 & 11.53 & 87.34 & 20.44 & 74.65 & 32.24 & 97.94 & 6.65 & 88.86 & 17.92 & 73.68 & 32.98 \\
\bottomrule
\end{tabular}
\vspace{-10pt}
\end{table*}

Three alternative routes fit this setting poorly. Speech enhancement front-ends have not been established as effective relative to models trained on noisy speech~\cite{yang2023}, and their processing artifacts, rather than residual noise, dominate downstream degradation~\cite{iwamoto2022,ochiai2024}; target speaker extraction~\cite{hsieh2026,ling2026} addresses competing talkers rather than additive noise. Audio-visual ASR~\cite{shi2022a,shi2022b,cappellazzo2025} followed by text matching reintroduces the lexical priors that QbyE avoids, which can mistranscribe the rare words and proper nouns that user-defined keywords may contain, and discards the continuous score on which a tunable operating point depends.

How should visual evidence enter such a system? A natural option is to let the visual stream predict a per-dimension mask and rescale the audio embedding with it, as in visually guided masking~\cite{xiong2025}, a strategy inherited from mask-based speech enhancement. Element-wise gating of this kind can reweight evidence already encoded in the audio representation, but cannot introduce visual information absent from it. We argue instead that the fusion stage must be able to \emph{inject} visual evidence into the audio representation, and that injection is only useful when the visual representation encodes phoneme-level distinctions in the first place. These two requirements are coupled: a mechanism with nothing to inject and a representation with no way in are equally inert.

We present GIVE-KWS\footnote{Code is available at \href{https://github.com/Padawan101/GIVE-KWS}{https://github.com/Padawan101/GIVE-KWS}.}, a QbyE-KWS system whose fusion stage, GIVE (Gated Injection of Visual Evidence), conditions the query audio on lip motion through gated cross-attention over a frozen self-supervised visual encoder. Our contributions are threefold. First, we show that a visual stream can be present yet inert, and link this inertness to the phonemic content of its representation. Second, crossing two visual encoders with two fusion mechanisms, we show that the two requirements above interact: the advantage of injection over masking widens with noise under a phoneme-bearing encoder and vanishes under a phoneme-poor one. Third, the resulting gain over the benchmark system grows with noise severity, reaching a 72.9\% EER reduction at $-10$~dB on unseen keywords.
\section{Proposed Method}
\label{sec:method}

\subsection{Framework overview}
\label{ssec:overview}
Let $w^{E}$, $s^{E}$ and $v^{E}$ denote the text, audio and video streams of an enrollment example, and $s^{Q}$, $v^{Q}$ the audio and video of a query segment. GIVE-KWS estimates
\begin{equation}
  p\!\left(s^{Q}, v^{Q} \mid w^{E}, s^{E}, v^{E}\right) \in [0,1],
  \label{eq:task}
\end{equation}
the probability that the query contains the same keyword as the enrollment, and is trained with binary cross-entropy. No component is indexed by keyword identity---the enrollment text is represented as a phoneme sequence and the decision is a comparison between representations---so the model applies unchanged to terms unseen during training.

Fig.~\ref{fig:arch} shows {GIVE-KWS}. Three frozen encoders produce the input representations. A grapheme-to-phoneme model\footnote{\href{https://github.com/Kyubyong/g2p}{https://github.com/Kyubyong/g2p}.} maps the enrollment keyword to a phoneme embedding sequence, Whisper-Tiny~\cite{radford2023} encodes audio at $50$~fps, and AV-HuBERT Base~\cite{shi2022a,shi2022b}, pretrained on LRS3~\cite{lrs3}, encodes the lip region at $25$~fps. All three are frozen and their outputs precomputed offline, so the trainable parameters reside entirely in the fusion, projection, alignment and matching stages.

The query audio representation first passes through the GIVE module (Sec.~\ref{ssec:fusion}), where visual evidence is injected. Each modality is then projected to a common $128$-dimensional space by a two-layer feed-forward network, after which a phoneme-level alignment stage (Sec.~\ref{ssec:phon}) anchors the audio and visual representations to the text phoneme embeddings.

Detection is performed by comparison. We pair each enrollment modality (text, audio and video) with each query modality (audio and video), producing six enrollment--query comparisons. For each pair, we compute a cosine similarity matrix $M \in \mathbb{R}^{T_{E} \times T_{Q}}$ between the two frame sequences. Each row of $M$ records how well one enrollment frame matches the query at every position; max-pooling along the query axis yields, for each enrollment frame, its best match anywhere in the query, which makes the comparison insensitive to where in the query the keyword occurs and to differences in speaking rate. The pooled vector is then used to attend over the full matrix, recovering the distribution of matches that pooling discards, and a recurrent layer summarizes the result into a fixed-size embedding. The six embeddings are concatenated and mapped to the detection score of (\ref{eq:task}) by a linear layer and a sigmoid. This matching design follows~\cite{xiong2025}, and we keep it unchanged so that performance differences are attributable to the components described next.

\subsection{Visual representation}
\label{ssec:visual}
A fusion module can only transfer information that the visual representation already encodes, so the encoder should be chosen for what it encodes. AV-HuBERT is a self-supervised audio-visual encoder whose masked-prediction pretraining over large unlabeled corpora encourages representations organized around speech units. We call a representation phoneme-bearing when frame-level phoneme labels are linearly recoverable from it; Sec.~\ref{ssec:diagnosis} compares AV-HuBERT with CNN-ResNet~\cite{zhou2022}, a task-trained convolutional encoder, on this basis.

\subsection{Gated injection of visual evidence}
\label{ssec:fusion}
The GIVE module conditions the query audio representation $\X$ on the query visual sequence $\V$. Audio frames act as queries and the visual sequence as keys and values in a multi-head cross-attention layer,
\begin{equation}
  \A = \mathrm{MHA}\!\left(\mathrm{LN}(\X),\; \mathrm{LN}(\V),\;
      \mathrm{LN}(\V)\right),
  \label{eq:attn}
\end{equation}
so that each audio frame retrieves the visual context most relevant to it. The attended evidence is then blended into the audio stream under a learnable scalar gate,
\begin{equation}
  \X' = (1-w)\,\X + w\,\A, \qquad w = \tanh(\alpha),
  \label{eq:fuse}
\end{equation}
followed by a feed-forward branch added residually under its own zero-initialized gate $\tanh(\alpha_{\mathrm{FFW}})$. The gating formulation follows the design used for visual conditioning of frozen language and speech models~\cite{rouditchenko2024,rouditchenko2025}.

Four properties of this design matter for our setting. First, $\A$ carries content absent from $\X$, which no rescaling of $\X$ can supply. Second, the interpolation in (\ref{eq:fuse}) trades the audio representation off against the attended visual context through a single learned weight shared across all inputs. Third, $\alpha$ is initialized to zero, so the module begins as an identity map and training starts from the audio-only solution. Fourth, cross-attention removes the need for explicit frame-rate alignment between the $50$~fps audio and $25$~fps visual streams.

The GIVE module is applied to the query branch only, since enrollment audio is clean by construction, and operates on the pre-projection audio representation so that the injected evidence is available to every downstream stage.

\subsection{Phoneme-level supervision}
\label{ssec:phon}
Because the evaluation vocabulary is disjoint from the training vocabulary, the representation must be organized around phonemes rather than around whole keywords. Two auxiliary objectives enforce this.

The alignment stage~\cite{xiong2025} randomly substitutes a fraction of frames between the audio and visual sequences, processes both streams with a shared self-attention block, and pools frames within forced-alignment phoneme boundaries under a fixed Gaussian kernel centered on each phoneme. Substitution encourages the shared block to produce modality-invariant representations, since a mixed sequence must remain interpretable; Gaussian pooling down-weights frames near boundaries, which are both less reliably aligned and more affected by coarticulation. An alignment loss $\L_{\mathrm{align}}$ then pulls the pooled audio and visual phoneme vectors towards the corresponding text phoneme embedding, which is unaffected by acoustic noise and therefore serves as a stable anchor.

A phoneme-level contrastive loss $\L_{\mathrm{cont}}$~\cite{li2025} operates on the pooled phoneme representations of the matching branches, pulling together representations of the same phoneme across modalities while pushing apart different phonemes; pairs sharing a phoneme label are excluded from the negative set so that the objective does not penalize genuine matches. Both objectives act only during training; inference requires no forced alignment.

\subsection{Training objective}
\label{ssec:loss}
The full training objective is
\begin{equation}
  \L = \L_{\text{det}}
     + \beta\,\L_{\mathrm{align}}
     + \gamma\,\L_{\mathrm{cont}},
  \label{eq:loss}
\end{equation}
where $\L_{\text{det}}$ is the binary cross-entropy on the detection score, $\L_{\mathrm{cont}}$ is summed over the six matching branches, and $\beta$ and $\gamma$ are given in Sec.~\ref{ssec:setup}.
\section{Experiments}
\label{sec:exp}

\subsection{Setup}
\label{ssec:setup}
\noindent\textbf{Dataset.} We use MISP-QEKS~\cite{xiong2025}, a tri-modal QbyE-KWS benchmark of 610,000 enrollment--query pairs derived from LRS2~\cite{lrs2}, with recorded environmental noise mixed at $5$, $0$, $-5$ and $-10$~dB. It provides two evaluation splits of 50,000 pairs each: Eval-seen, whose keywords appear in the training vocabulary, and Eval-blind, whose keywords do not. Speakers in both evaluation splits are disjoint from the training set~\cite{xiong2025}. Its reference system, XEQ-Matcher~\cite{xiong2025}, is the benchmark system throughout.

\noindent\textbf{Implementation.} Following the benchmark recipe~\cite{xiong2025}, we train with SGD at an initial learning rate of $0.01$ with step decay and a batch size of $128$, apply noise to $80\%$ of training samples at SNRs drawn from $\{-10, -5, 0, 3, 6, 9\}$~dB, and set $\beta = 0.5$ and $\gamma = 0.1$. These values are shared by every system and were not tuned. Our retrained XEQ-Matcher matches the results reported in~\cite{xiong2025}. GIVE uses a single cross-attention layer with six heads of 64 dimensions and a feed-forward expansion of four, following standard Transformer practice. Training is performed on a single NVIDIA RTX 4090 GPU and takes approximately 72 hours.

\noindent\textbf{Evaluation.} We report ROC-AUC and EER on the clean condition and four noisy conditions ($5$, $0$, $-5$, $-10$~dB), summarized as three scenarios: the clean condition; the unweighted average over all five conditions, following the benchmark; and $-10$~dB, the extreme case in which the acoustic evidence is largely buried and visual input matters most. All audio-visual systems share the auxiliary objectives of Sec.~\ref{ssec:phon} and differ only in the visual encoder and fusion stage; the text\,+\,audio system omits the alignment objective and the contrastive terms involving vision. All systems are evaluated after three training epochs.

\noindent\textbf{Significance.} Because EER aggregates two error rates, standard tests for proportions do not apply directly~\cite{bengio2004}; differences are tested with a paired bootstrap~\cite{bisani2004} over enrollment--query pairs (10,000 resamples); each pair is resampled together with all of its SNR conditions~\cite{liu2020}, and the EER difference is recomputed on every resample. Comparisons supporting a claimed improvement use a Bonferroni-corrected level of $0.05/6$ (three scenarios on each split); comparisons supporting an absence of effect use the uncorrected $95\%$ interval. Each system is trained once, so the intervals reflect sampling of evaluation pairs rather than training variance.

\subsection{Is the visual stream used?}
\label{ssec:diagnosis}
We first ask whether a task-trained visual encoder contributes to detection. A system trained with text and audio only gives $35.15\%$ EER at $-10$~dB on Eval-blind (Table~\ref{tab:main}), against $33.19\%$ with the CNN-ResNet visual stream; on Eval-seen the two are indistinguishable ($33.53\%$ vs $33.52\%$). The two systems thus stay within 2.0 percentage points of each other at $-10$~dB, where visual input is expected to matter most.

Linear probing points to the representation. We train a linear classifier to predict the forced-alignment phoneme label of each frame from frozen features under an identical budget for every encoder, and report its accuracy over the first 2,000 batches, scoring each batch before updating on it. The task-trained CNN-ResNet reaches $6.2\%$ frame-level phoneme accuracy and $19.6\%$ viseme accuracy, against $27.0\%$ and $41.2\%$ for AV-HuBERT\@. The audio branch reaches about $58\%$ phoneme accuracy for reference. Sec.~\ref{ssec:main} shows that repairing the representation is nonetheless not sufficient on its own.

\subsection{Main results}
\label{ssec:main}
Relative to the benchmark system, the gain of GIVE-KWS grows with noise severity on both splits (Table~\ref{tab:main}), reaching a $72.9\%$ EER reduction at $-10$~dB on Eval-blind.

The four middle rows of Table~\ref{tab:main} separate the two factors. With the fusion mechanism held fixed to masking, replacing the visual encoder lowers EER at $-10$~dB from $33.19\%$ to $10.14\%$ on Eval-blind: what the representation encodes matters. With the encoder held fixed to AV-HuBERT, replacing masking by injection lowers it further, by $11.1\%$ relative on Eval-blind and $25.2\%$ on Eval-seen, both significant: how it enters the audio stream matters as well.

\begin{figure}[t]
\centering
\includegraphics[width=0.95\columnwidth]{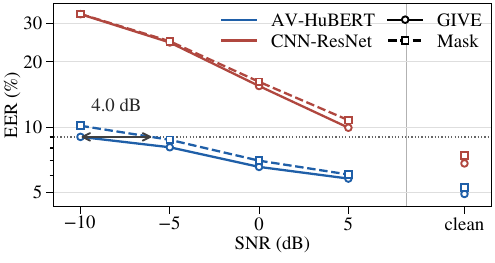}
\vspace{-15pt}
\caption{EER on Eval-blind as a function of SNR\@. Color denotes the visual encoder and line style the fusion mechanism. As noise increases, injection and masking diverge under AV-HuBERT but converge under CNN-ResNet. The arrow marks the effective SNR gain of injection with AV-HuBERT at $-10$~dB. The dotted line marks the EER of GIVE-KWS at $-10$~dB, which the benchmark system (red, dashed) does not reach at any noisy SNR.}
\label{fig:snr}
\vspace{-10pt}
\end{figure}

Fig.~\ref{fig:snr} expresses the fusion effect as an effective SNR gain~\cite{linharte}, the shift in SNR at which the weaker system matches the stronger one's EER at $-10$~dB, obtained by linear interpolation between tested SNRs: under the same encoder, injection yields a gain of $4.0$~dB over masking on Eval-blind (95\% CI $[2.9, 5.1]$) and $9.3$~dB on Eval-seen ($[8.5, 10.1]$), and against the benchmark system the gain exceeds the $15$~dB we can measure.

The two conditions interact rather than add (Table~\ref{tab:main}, Fig.~\ref{fig:snr}). With AV-HuBERT, the advantage of injection grows with noise and is significant in five of six conditions; the exception is the clean condition on Eval-blind, where the acoustic evidence is intact and vision has little to add. With CNN-ResNet the advantage all but vanishes at $-10$~dB, where the $95\%$ intervals include zero on both splits. In absolute terms, injection lowers EER at $-10$~dB by $4.71$ versus $0.34$ points on Eval-seen and $1.13$ versus $0.06$ on Eval-blind, more than an order of magnitude apart and with non-overlapping confidence intervals on both splits.

All systems perform worse on Eval-seen than on Eval-blind. The two splits differ in more than vocabulary overlap: they are drawn from different speakers, and Eval-seen keywords are significantly shorter ($7.31$ vs $7.67$ phonemes on average; $p<0.001$, Mann--Whitney U), leaving fewer phonetic cues per trial. The two splits should therefore not be read as a controlled in-vocabulary versus out-of-vocabulary comparison; the conclusions above hold on both.

\subsection{Ablations}
\label{ssec:ablation}
Removing the visual stream from GIVE-KWS at inference raises EER at $-10$~dB from $9.01\%$ to $32.91\%$ on Eval-blind and from $13.96\%$ to $32.21\%$ on Eval-seen, comparable to the systems with CNN-ResNet ($33.19\%$ and $33.52\%$). Replacing the visual encoder output with zeros while keeping the pathway active gives $32.98\%$ and $32.24\%$, indistinguishable from removal, indicating that the gain depends on the visual content rather than merely on the fusion pathway and its parameters. Whereas the CNN-ResNet visual stream contributes little (Sec.~\ref{ssec:diagnosis}), the low-SNR gain of GIVE-KWS relies on informative visual features.
\section{Conclusion}
\label{sec:conclusion}

We presented GIVE-KWS, a query-by-example keyword spotting system whose fusion stage injects visual evidence through gated cross-attention, and used it to show that visual robustness depends on two interacting conditions: a visual representation carrying phonemic information, and a fusion mechanism able to inject that information rather than rescale audio features. With a phoneme-poor encoder, injection offers no measurable benefit at $-10$~dB. With a phoneme-bearing one, its advantage over masking grows as the acoustic evidence degrades. Since the visual representation remains less linearly discriminative of phonemes than the audio branch, improving it---for instance through lip-reading-adapted or viseme-guided pretraining---is a promising direction alongside further study of fusion.

\vfill\pagebreak



\begingroup
\hbadness=2000
\bibliographystyle{IEEEbib}
\bibliography{refs_2}
\endgroup

\end{document}